\documentclass[10pt, conference, compsocconf]{IEEEtran}

\pdfoutput=1 

\usepackage{amsmath,epsfig, txfonts, amssymb, mathrsfs}
\usepackage{cite} 
\usepackage{url}

\DeclareMathOperator*{\argmin}{argmin}

\begin{document}

\title{A sparse regulatory network of copy-number driven expression reveals putative breast cancer oncogenes}
\author{\IEEEauthorblockN{Yinyin Yuan, Christina Curtis, Carlos Caldas, Florian Markowetz}
\IEEEauthorblockA{Cancer Research UK, Cambridge Research Institute, UK\\
Department of Oncology, University of Cambridge, UK\\
Email: yy341@cam.ac.uk}
}

\maketitle
\IEEEpeerreviewmaketitle
\begin{abstract}
The influence of DNA {\it cis}-regulatory elements on a gene's expression has been intensively studied. However, little is known about expressions driven by  {\it trans-}acting DNA hotspots. DNA hotspots harboring copy number aberrations are recognized to be important in cancer as they influence multiple genes on a global scale. The challenge in detecting {\it trans-}effects is  mainly due to the computational difficulty in detecting weak and sparse {\it trans}-acting signals amidst co-occuring passenger events. We propose an integrative approach to learn a sparse interaction network of DNA copy-number regions with their downstream targets in a breast cancer dataset. Information from this network helps distinguish copy-number driven from copy-number independent expression changes on a global scale. Our result further delineates {\it cis-} and {\it trans-}effects in a breast cancer dataset, for which important oncogenes such as ESR1 and ERBB2 appear to be highly copy-number dependent. Further, our model is shown to be efficient and in terms of goodness of fit no worse than other state-of the art predictors and network reconstruction models using both simulated and real data.
\end{abstract}

\section{Introduction}

\begin{figure}[tb]
\centering
\includegraphics[width=.5\textwidth]{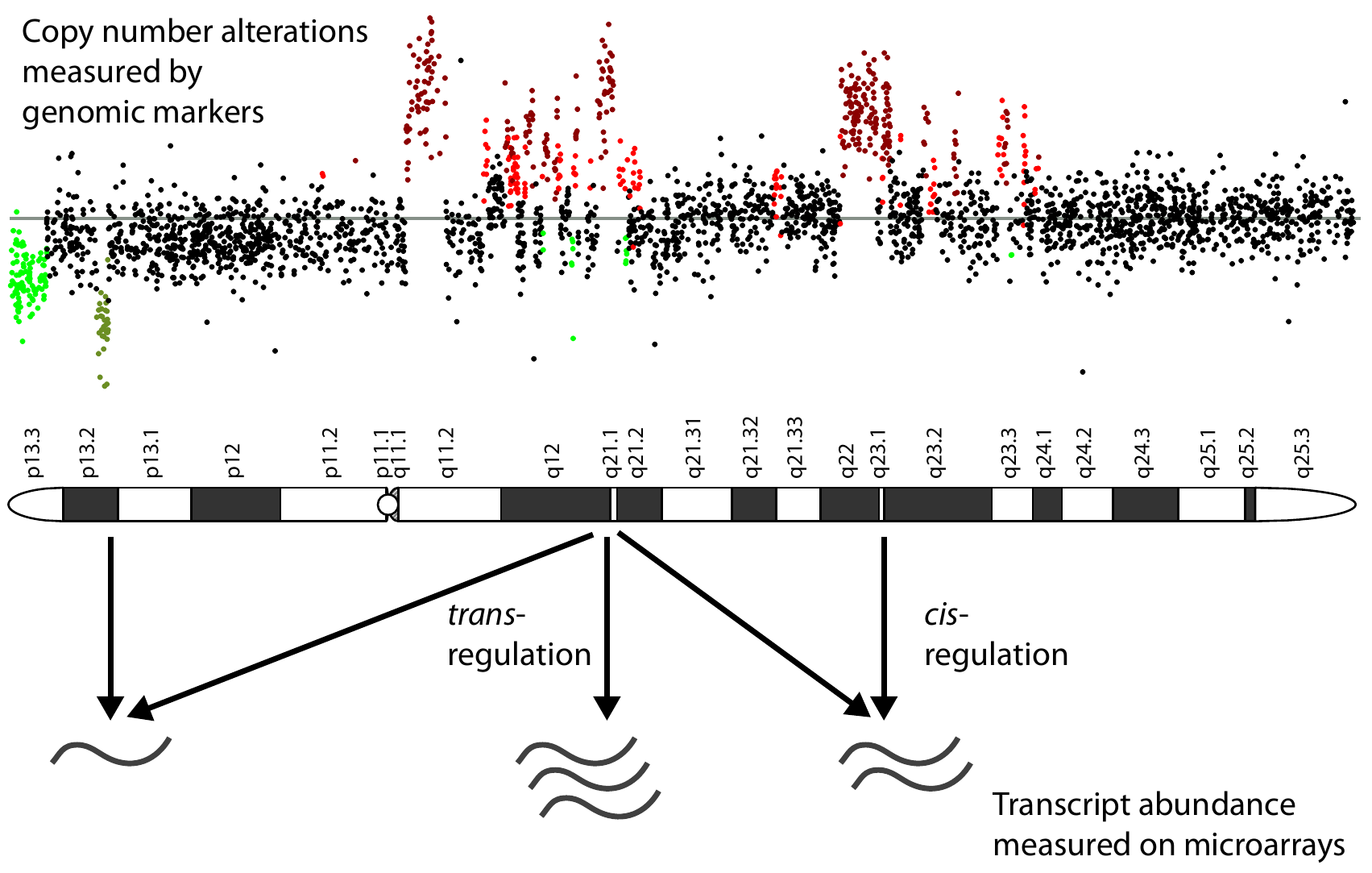}
\caption{ Expression abundance level can be affected by both {\it cis}- and {\it trans}- acting copy number alterations. integrating gene expression data and gene dosage information measured by microarray or sequencing technologies. }
\label{dna-mrna}
\end{figure}

Copy number alterations, including both germline variants (CNVs) and somatic copy number aberrations (CNAs) are associated with disease \cite{pollack99genomewide}. Somatic aberrations (CNAs) are particularly important for tumourigenesis, contributing to genomic instability and gene deregulation. For example, oncogene activation by gene amplification or tumour suppressor loss as a result of gene deletion (Fig.\ref{dna-mrna}) can cause transcriptional changes and contribute to the pathogenesis of cancer. 
On the other hand, expression can be influenced by the proximal genes within a several Mb window (\emph{cis}-acting), but can also be affected by a remote CNA throughout the genome (\emph{trans}-acting), as depicted in Fig.\ref{dna-mrna}. One of the challenges in cancer genomics is to characterize such intermediate phenotypic changes caused by both {\it cis-} and {\it trans-} CNAs that ultimately lead to tumour progression.

\paragraph{Detecting {\it cis}- and {\it trans}-regulatory effects}
An integrative analysis of CNA and expression has the potential to uncover regulatory relationships between CNA and gene expression (Fig.\ref{dna-mrna}). The availability of genome-wide copy number and gene expression data facilitates the discovery of such regulatory maps, but few studies have quantitatively assayed the effect of both {\it cis-} and {\it trans-} copy number on the expression abundance in the same primary tumour sample. Given the inherent noise in such data and the large-scale of genomic information, the detection of both {\it cis}- and {\it trans}-acting effects between DNA copy number and mRNA on a genome-wide basis is a difficult task.  Such effects are small and infrequent \cite{sladek06elucidating}, which further imposes computational challenges. Detection of  the driving alterations amidst numerous random passenger events is complicated by the frequency of co-occurring events such as co-amplifications and co-deletions. Thus despite considerable efforts, the location of many relevant CNAs that result in gene disregulation remain elusive.

\paragraph{Previous research}
This area of research is of key importance since many somatic alterations could contribute to tumorigenesis and their identities are critical for the development of specific anti-cancer agents. Not surprisingly, the association between DNA copy number and mRNA expression data has gained considerable attention. One of the first papers to combine gene expression and DNA copy number data by  \cite{pollack99genomewide} concluded that 62\% of highly amplified genes demonstrated moderate to high expression levels. Others have similarly identified genes that exhibit gain/loss and simultaneous over/under-expression in cancer \cite{myllykangas08integrated}, but it has been noted that chromosomal amplifications do not necessarily result in the over-expression of the genes located in the same cytoband  \cite{platzer02silence}.

In summary, recent studies have focused on the local effects of copy number on expression, and few have considered {\it trans}-acting effects on a global scale. Correlation analysis remains in widespread use for these types of comparisons due to its efficiency \cite{bussey06integrating}, and has more recently been used to query {\it trans}-effects on a genome-wide level \cite{lee08integrative}.  Several more sophisticated tools have also been proposed.  For example,  \cite{menezes09integrated} made use of gene sets instead of individual genes in regression models in their efforts to integrate the two data types. However, simple regression has limited power in face of the immense amount of genomic data. Moreover, their focus is on {\it cis}- events and therefore cannot incorporate remote interactions.  

\paragraph{Our approach}
In order to find copy-number driven expression, we propose a new framework for reconstructing a genome-wide regulatory map on a genomic and transcriptomic level. The objective of the proposed framework is to set up an interaction network between CNA and expression changes (arrows in Fig.\ref{dna-mrna}). By interpreting global gene expression changes in the context of CNA, the network has the potential to uncover both {\it cis}- and {\it trans}-regulatory elements.
Our approach combines several key ideas: First, we employ an $L_1$-regularized regression model which allows simultaneous feature selection and model fitting which generalizes well in high-dimensions. Second, we introduce a combination of global and local search strategies to optimally set the regularization parameter in our model. Third, we quantitatively measure each transcript's  copy-number dependence. 


\section{Method}
We assume copy number data of $p$ DNA regions $X= (x_1, x_2, ..., x_p)$  and mRNA expression data of $q$ genes $Y= (y_1, y_2, ..., y_q)$, both available from $n$ tumour samples.

The framework for investigating CN trans- effect is built on a linear regression method with $L_1$-regularization (section~\ref{method-lasso}). The strength of regularization is chosen adaptively for each gene by a combined global and local search over the space of regularization parameters (Section~\ref{method-search}). The results are incorporated into a quantitative measure for copy-number dependence by comparing variances of the model residuals with and without a locus of interest (Section~\ref{method-granger}).

\subsection{$L_1$-constrained regression selects most influential genomic loci for each transcript}\label{method-lasso}
We use $L_1$ constrained regression (a.k.a the {\it Lasso} \cite{lasso}) to derive the causal effects of genome-wide copy number data, as the predictor variables, on gene expression profiles, as the response variables. $L-1$-regression minimizes the sum of squared errors between response and prediction, while keeping a bound on the sum of the absolute values of the regression coefficients. By penalizing the absolute sum of coefficients, $L_1$-constrained regression effectively shrinks non-significant coefficients to zero. In contrast to this, $L_2$-constrained regression (aka \emph{ridge regression}) keeps a bound on the sum of the squared values of the regression coefficients and generally results in small, but non-zero, regression coefficients. Hence,  $L_1$-regression is a prominent method in high-dimensional studies for its efficiency in performing sparse statistical inference on thousands of variables \cite{tibshirani97lasso}. Efficient computational methods exist to infer the entire path of variable inclusions over different choices of regularization strength \cite{lars}.

\paragraph{In our scenario:} For gene $i$  with expression profile $y_i$, let $X \beta_i$ be the candidate prediction model with regression coefficients $\beta_i = (\beta_{i1}, \beta_{i2}, ..., \beta_{in})$. The Lasso model is then given by  
\begin{equation}\label{EqLasso}
\hat{\beta}_i = \argmin_{\beta_i} \   \left\|\  y_i - X \beta_i \  \right \|^2_2 +\lambda_i  \left\| \beta_{i} \right\|_1 ,
\end{equation}
where $\| x \|_2 = \sum x_i^2$ indicates the $L_2$ norm measuring the distance between model prediction, and response and $\| x \|_1 = \sum |x_i|$ indicates the $L_1$ norm used to constrain the regression coefficients.

\begin{figure}[t]
\centering
\includegraphics[width=.48\textwidth]{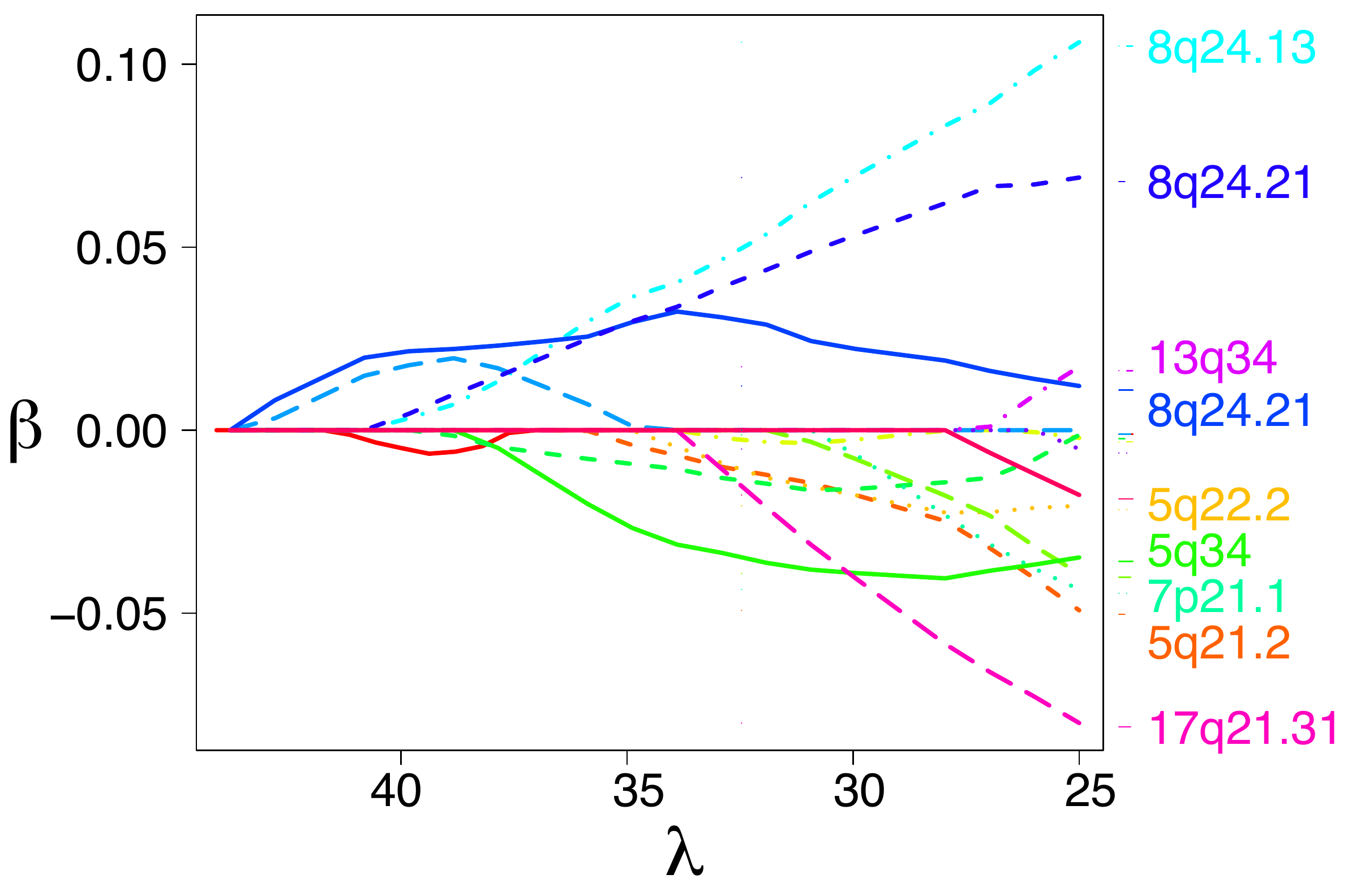}
\caption{The  Lasso path along a range of $\lambda$ values for  {\it MYC}, a well known oncogene,  in the experimental dataset. Each line corresponds to a regression coefficient measuring the influence of copy number variation at a particular locus (names on the right) on {\it MYC} expression. Regularization strength decreases with $\lambda$ from the left to the right. This can be seen by more and more coefficients hitting the zero line and vanishing from the model.}
\label{lambda}
\end{figure}

The $\lambda_i$ in Eq.(\ref{EqLasso})  is a regularization parameter controlling the sparsity and  strength of regularization. 
By varying $\lambda_i$,  $L_1$-regression/Lasso adaptively includes predictors in the model (see Figure~\ref{lambda}). The subscript ``$i$" emphasizes that the optimization problem in Eq.(\ref{EqLasso}) has to be solved for every gene. In particular, for each gene a regularization parameter $\lambda_i$ has to be chosen. In the next section, we propose to utilize an efficient combination of a global and local search strategy.

\subsection{Adaptive regularization by combined global and local search}\label{method-search}

\begin{figure}[!h]
\centering
\includegraphics[width=.45\textwidth]{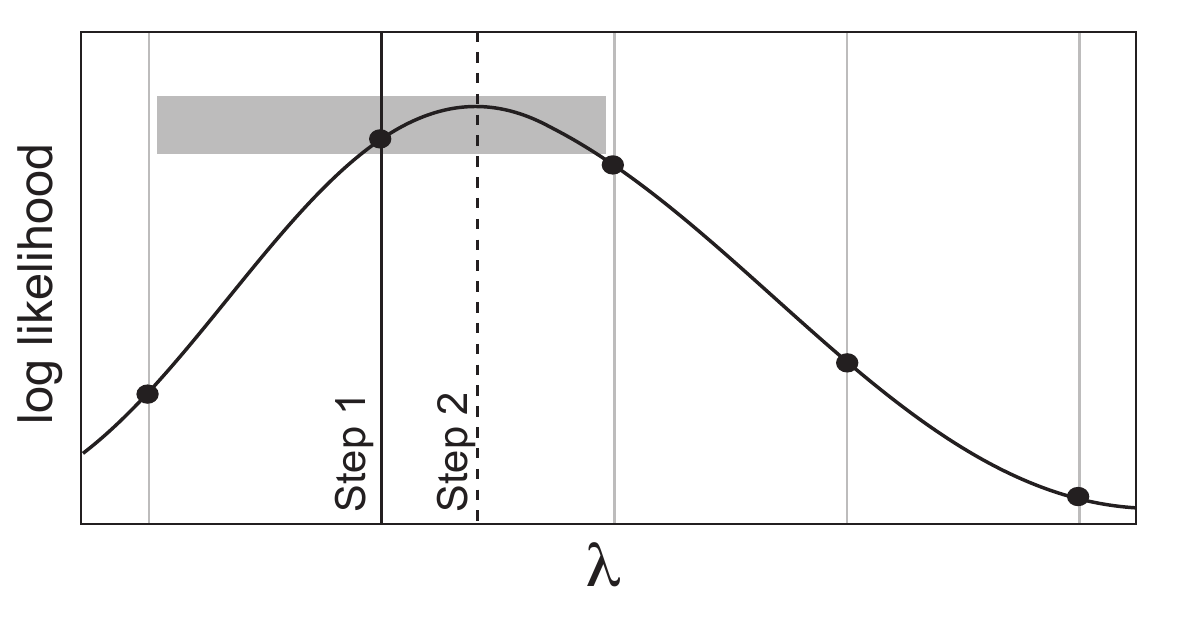} 
\caption{$\lambda$ selection procedure base on the log-likelihood function. The grey lines mark the global search points, and the two black lines (one solid, one dashed) mark the optimal selection points of $\lambda$ by Step 1 and 2.}
\label{loglike}
\end{figure}

The regularization parameter $\lambda$ is crucial since it directly determines how many predictor vectors are to be included into the model. 
We perform a two-step procedure for $\lambda$ selection that combines a global search over a discrete set of candidate values with a local search around the optimal candidate.
For the global search the likelihood function is evaluated by cross-validation, for the local search Brent's minimization without derivatives \cite{brent73} implemented in R package ``penalized'' \cite{geoman09l1} is used. 
Brent's method combines steps of golden section (a fast method based on binary divisions of the search space) with parabolic interpolation to efficiently zoom in onto the maximum.
\begin{enumerate}
\item[Step 1] Global search: We perform 10-fold cross-validation across a wide range of candidate $\lambda$ values (grey lines in Fig.\ref{loglike}). Log-likelihood scores (black dots) are computed for each of these values to provide a point of focus for the next step.
\item[Step 2] Local search: At the point of the maximum log-likelihood of Step 1 (black line in Fig.\ref{loglike}), we perform a local search (grey box) with Brent's optimization to locate the optimal value of $\lambda$ (dashed line).
\end{enumerate}

Having chosen an optimal $\lambda$ value and computing the Lasso solution of Eq.(\ref{EqLasso}) for each gene, we obtain a coefficient matrix $B = (\hat{\beta}_1, \hat{\beta}_2, \ldots, \hat{\beta}_n)$. Because of the $L_1$-constraint in the Lasso, this matrix will be sparsely populated and many coefficients will be zero. Matrix $B$ represents the regulatory relationships  between copy number changes and transcriptional changes that was indicated by arrows in Fig.\ref{dna-mrna}.

\paragraph{A technical note} Lasso assumes the predictor variables to be linearly independent. However, DNA copy number data obtained from high resolution microarrays can exhibit high correlation between probes. Therefore, copy number data should be first tested for independence and, if necessary, the data should be merged with an appropriate method such as those implemented in R package ``CGHregion'' \cite{cghregions} to produce regions with independent signatures. We implemented our framework in R with the packages ``penalized"  and ``CGHregions".

\begin{figure}[!b]
 \sffamily
 \begin{minipage}{.48\textwidth}
\textbf{A.}  Precision curves 

\centerline{\includegraphics[width=.95\textwidth]{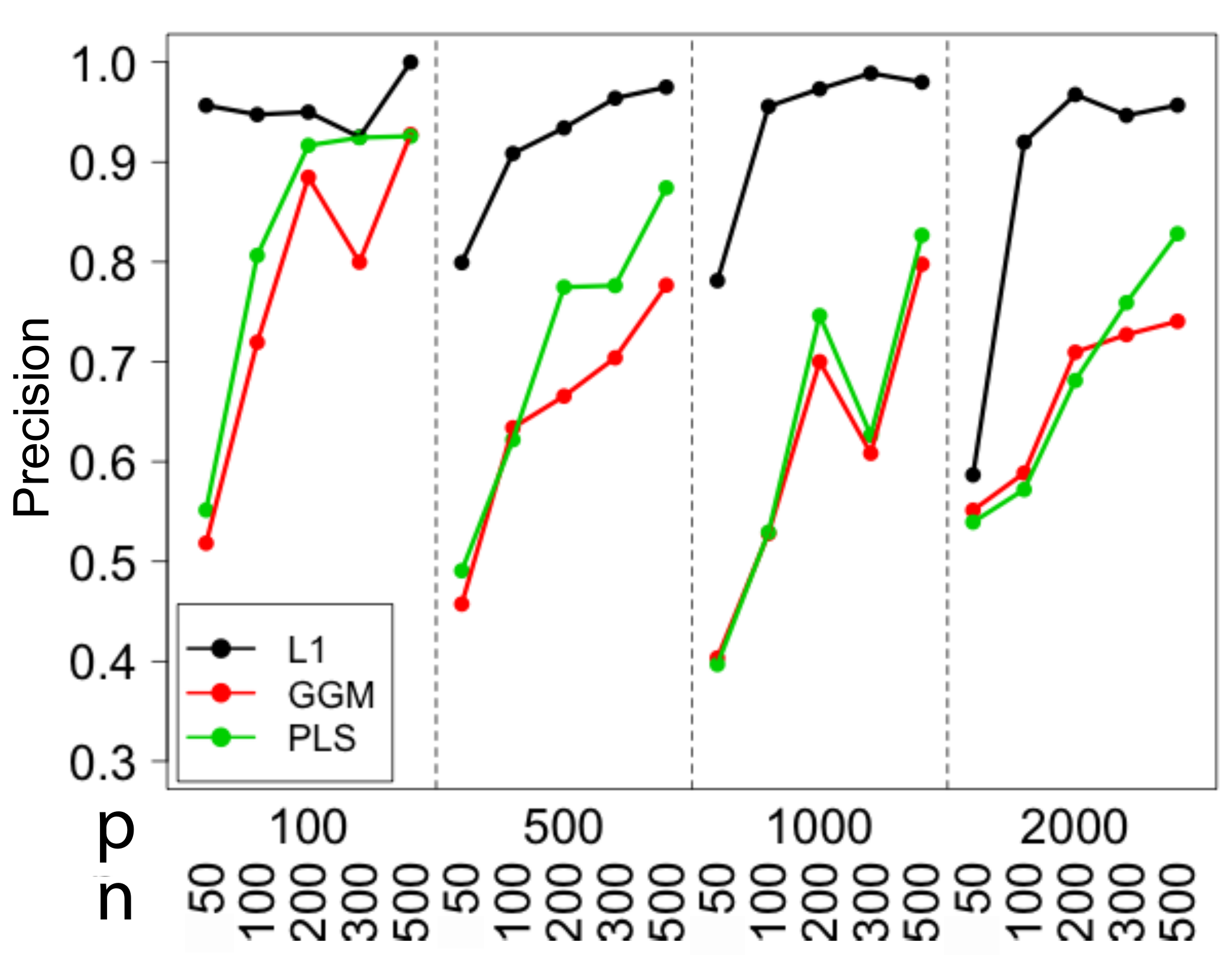}}
\end{minipage}
\hfill
 \begin{minipage}{.48\textwidth}
\textbf{B.} Recall curves 

\centerline{ \includegraphics[width=.95\textwidth]{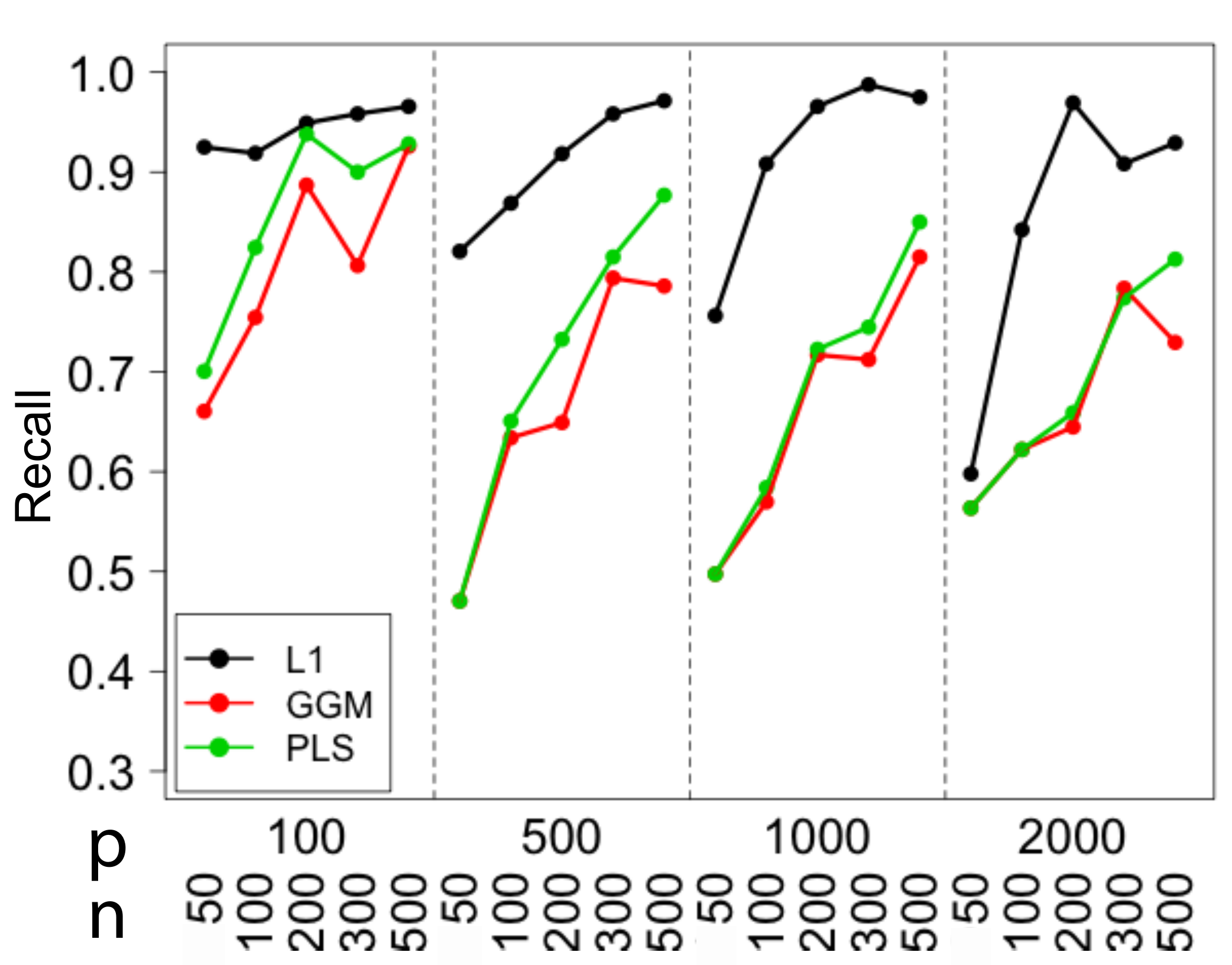}}
\end{minipage}

\caption{Performance of three methods in the experiments of network structure learning, data is simulated with varied number of samples $n$ and number of predictor variables $p$ and number of response variables $q, p=q$. }
    \label{structure}
\end{figure}

\subsection{A quantitative measure of copy-number dependence}\label{method-granger}


The proposed framework allows us to quantitatively  measure the dependency between gene expression and copy number, both in {\it cis} and in {\it trans}.  The underlying assumption is that the predictive power of a model can be measured by the proportion of variance explained, i.e., the ratio of the variance of modelling residuals with or without incorporating  one or more predictors.  More formally, given two models --one {\it with} and one {\it without} incorporating a set of predictors-- the negative natural logarithm of the variance ratio between the prediction residuals,
\begin{equation}\label{EqGranger}
\text{score}_i \ = \ - \ln \left(\frac{\sigma^2_{with}}{\sigma^2_{without}}\right),
\end{equation}
can be used as a score indicating the dependency of the response variable on the predictor(s). 
Therefore, by checking if the variance of residuals is reduced after incorporating the copy-number predictors,  the dependency of expression responses on copy number can be quantitatively measured. 

Furthermore, comparisons can be conducted with copy-number {\it cis}- and {\it trans}-, indicating whether a gene is significantly influenced by the {\it trans}- effect.
Formally, we do this by comparing a sequence of nested models that successively incorporate more and more copy number predictors: from no predictor at all, to using only copy number changes in  {\it cis}, and finally to using both  {\it cis}- and  {\it trans}- copy number changes. 
\begin{eqnarray}
y_i &=&y_i \label{EqPrediction1}\\
y_i &= &x_i\beta_i  + \epsilon_i \label{EqPrediction2}\\
y_i &=&x_i\beta_i  + \sum_{j, j \neq i} x_j \beta_j + \epsilon_i' \label{EqPrediction3}
\end{eqnarray}
Eq.(\ref{EqGranger}) then allows us to estimate the effect of adding more predictors by measuring the difference of variance explained between any two equations in Eq.(\ref{EqPrediction1})-(\ref{EqPrediction3}).
For example, the dependency score of  expression of a gene $i$' on copy number can be calculated based on Eq.(\ref{EqPrediction1}) and Eq.(\ref{EqPrediction3}) 
\begin{equation}\label{EqCopyNumberDependent}
\text{score}_i^{both} \ = \ -\ln \left(\frac{\sigma^2_{\epsilon_i'}}{\sigma^2_{y_i}}\right).
\end{equation}
And the  dependency score of expression $y_i$ on the copy number {\it trans}-effect conditioned on the {\it cis}- effect based on Eq.(\ref{EqPrediction2}) and Eq.(\ref{EqPrediction3}) is 
\begin{equation}\label{EqTransDependent}
\text{score}_i^{trans} \ = \ -\ln \left(\frac{\sigma^2_{\epsilon_i'}}{\sigma^2_{\epsilon_i}}\right).
\end{equation}
If the score is significant, there is likely to be a copy number {\it trans}-effect on gene $i$. Note that by incorporating the  {\it cis}- predictor $x_i$ in both equation, we remove the {\it cis}-effect before exploring the {\it trans}-effect $ \sum_{j, j \neq i} x_j$, which is equivalent to a conditional independence test.


\section{Experiments on simulated networks} \label{simulation}

Simulated data was generated to test the performance of the $L_1$ method on network structure inference. Let $X\sim N(0, 1)$ be a matrix with $p$ Gaussian variables each of $n$ samples, and $T$ be a sparse matrix of $p$ by $q$ generated from randomly sampling 50 non-zero coefficients from a distribution $N(0, 1)$. From $X$ and $T$, a response vector $Y = XT$  with $q$ variables is generated. Both $X$ and $Y$ are added with centered Gaussian noise $N(0, \sigma^2_\epsilon)$. The objective is to infer $T$ from $X$ and $Y$.

Two popular network reconstruction methods in the literature that are applicable to the small $n$, large $p$ problem are compared with the $L_1$ method. They are the Gaussian Graphical Models  \cite{schafer05empirical} and Partial Least Square \cite{vasyl08reconstruction}. However, to remove any ambiguity that could be caused by the different numbers of inferred relations, the number of relations, i.e. total number of non-zero coefficients in $T$ are known to all methods.

The performance of the methods in comparison is judged by whether the non-zero coefficients in $T$ can be correctly detected. Data is simulated with $p$, the number of variables in $X$,  varying from 100 to 2000 and sample size $n$ varying from 50 to 500. The number $q$ of variables in $Y$ is equal to $q$. For each setting of $p$ and $n$, datasets are generated with different noise levels $\sigma^2_\epsilon \in \{0.05, 0.1, 0.2\}$. For the result, precision and recall statistics for the inference results of all variables in $Y$ are computed and averaged across noise levels. 

The precision and recall curves, grouped by values of $p$, are plotted in Fig.\ref{structure}. The advantage of the $L_1$ method can be clearly seen:   in all scenarios precision and recall are higher than for the competitors. The difference is most pronounced for large values of $p$, which is especially important for real-world applications. In summary the simulations show that the proposed method outperforms other state-of-the-art methods in its accuracy to reconstruct regulatory networks. 

%
%
%
%

\section{A genome-wide map of $cis$- and $trans$-regulation in breast cancer} \label{results}

To demonstrate the effectiveness of the proposed method in a real-world application we used a breast cancer dataset with 89 samples analyzed using both mRNA expression and DNA copy number microarrays \cite{chin06genomic}.  

\subsection{A quantitative assessment of prediction accuracy} \label{experiment}

While the simulations showed the accuracy of our framework in network reconstruction compared to other network models,  on real data we can not repeat this analysis since the true network is unknown. However, a complementary question that we can answer is how well our model predicts gene expression from copy-number alterations compared to other state-of-the-art regression models. We compare $L_1$-constrained regression with three other well known prediction models: Random Forests \cite{randomforests}, SVM \cite{svm}, and Recursive Partitioning \cite{recursivepartitioning}. Each method uses copy number data to predict expression data and accuracy is measured in terms of the mean squared error of the prediction. 

Models were trained with a subset of samples in the \cite{chin06genomic} data and then the fitted models were used to predict the rest of samples.  In detail: 100 expression profiles are randomly selected from the Chin data. For each of the expression profile, models were fitted with 7/8 (78) of the samples and then the mean squared errors of predicting the test set 1/8 (11) of the samples. 
The Random Forests method was trained with 500 trees so that the computational time is on the same scale as that of other methods. SVM was trained with a linear kernel. 

\begin{figure}
\includegraphics[width=.45\textwidth]{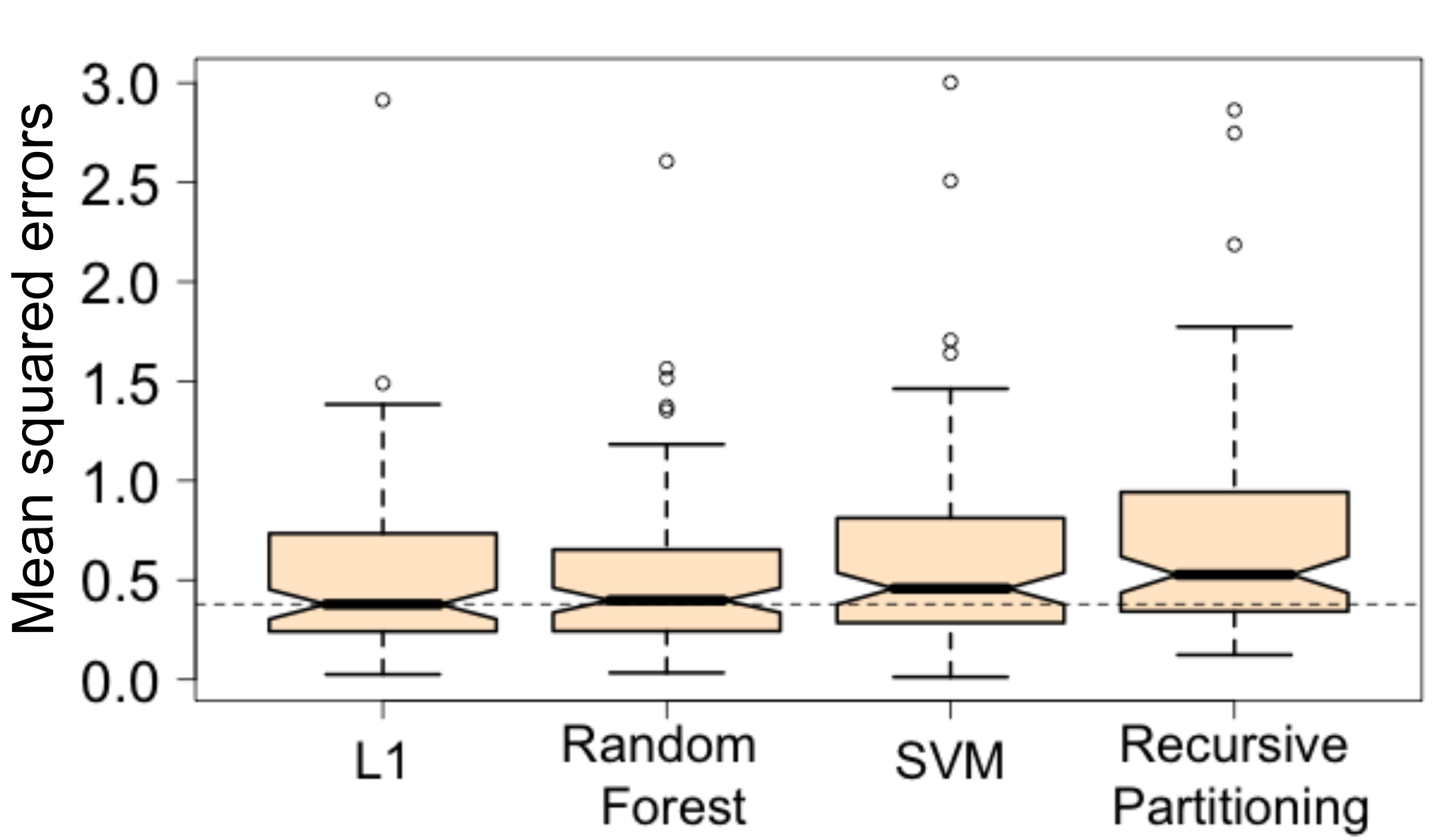} 
\caption{Mean squared errors of predictions by four methods on Chin's dataset, the dashed line shows the median of mean squared errors of Lasso. The average of the mean squared errors for four methods are 0.526, 0.528, 0.635, 0.707. }
\label{Fig.predictive}
\end{figure}

Results in Fig.\ref{Fig.predictive} show that the $L_1$-constrained regression is not out-performed by any competitor. Additionally, it provides a sparser solution than the other models: On average $L_1$-regression used less than 7 features, while Recursive Partitioning used more than 32 and SVM and Random Forests use all of them.  Thus, the $L_1$-model combines sparsity with high prediction accuracy. This is important since the selected features directly point to the relevant copy number alterations leading to expression changes. This experiment proves the proposed framework's potential with applications on copy number and expression data.

\subsection{The global impact of copy-number change on gene expression}

The proposed framework was used to study the impact of copy-number aberrations on every gene expression profile in the Chin dataset. 
Using Eq.(\ref{EqCopyNumberDependent}) we can rank genes based on the copy-number dependent score, as shown in Table~\ref{TableDependentGenes}. 
Several known oncogenes are present such as {\it ERBB2} and its neighbors, {\it STARD1} and {\it MGC9753}, as well as {\it ESR1}. For example, {\it ERBB2} is amplified in approximately 20\% of the breast cancer cases for which concomitant over-expression is also typically observed. Many other genes with less well characterized functions were also identified and merit additional investigation, e.g. TFF1 which is a stable secretory protein expressed in the gastric mucosa. 



\begin{table}[!htp]
\caption{Top 20 genes ranked by the copy-number dependent score}
\centering
\begin{tabular}{|c|c|c||c|c|c|} \hline
Gene &Chr  &  Score & Gene &Chr &  Score \\ \hline
TFF1 & 21 &   3.31 &MGC9753 & 17 &   2.38 \\
NAT1 &  8 &  2.73  &PROML1 &  4 &  2.33\\
TFF3 & 21 &   2.63 &CEACAM6 & 19 &  2.31\\
AGR2 &  7 &   2.61 &SCGB2A2 & 11 &  2.31\\
GABRP &  5 &  2.53 &CPB1 &  3 & 2.27 \\
MGC9753 & 17 &   2.5 &MLPH &  2 & 2.26\\
ESR1 &  6 &  2.49 &GRB7 & 17 &   2.26 \\
S100A7 &  1 &  2.44 &SFRP1 &  8 &  2.26 \\
PIP &  7 &  2.39 &ERBB2 & 17 &   2.24\\
HNF3A & 14 &  2.38 &CEACAM5 & 19 & 2.21\\
\hline
\end{tabular}
\label{TableDependentGenes}\end{table}

Expressions driven by the same CNAs region are also enriched with cancer-specific functions.  We test whether the downstream targets of these DNA regions share similar functions or motifs, since genes with similar functions are often co-regulated. Gene Ontology (GO) \cite{GO} enrichment analysis was performed on the downstream targets for CNAs regions driving many expressions. 
Numerous GO terms with low $p$-values, indicative of a high degree of functional similarity, were identified. For example, the targets of 22q13.33 are significantly enriched for GO terms involving the immune response ($p$-value 6E-11), lymphocyte activation (9E-11), and leukocyte activation (5E-10), with the former two pertaining to the adaptive immune response. 

Genetic variation at 8q24 has been associated with both breast and prostate cancer \cite{meyer09genetic} and this region is frequently amplified in breast cancer. One of the top ranking hot spots, 8q24.13, has downstream targets enriched for GO terms such as the positive regulation of ubiquitin-protein ligase activity during mitotic cell cycle, and the proteasomal protein catabolic process. Promoter analysis revealed a significantly enriched unknown motif CGGCGACATA (hypergeometric test $p$-value 4.7E-19, FDR on 50 random samples $p$-value 0.0044) in the upstream (-2000 bp) region.  Another hot spot, 9q22.2, exhibits {\it trans}-effects on {\it TOP2A} (17q21) and the MCM family, and has downstream targets enriched for DNA replication and mitotic cell cycle functions. Once again, these findings are in-line with existing knowledge on cancer biology amassed using various approaches.


\subsection{Individual example of copy-number driven expression}

\begin{figure}[t]
\centering
 \sffamily
\textbf{A.}  Correlation analysis of {\it MYC} expression with genome-wide copy number data
  
\centerline{\includegraphics[width=.5\textwidth]{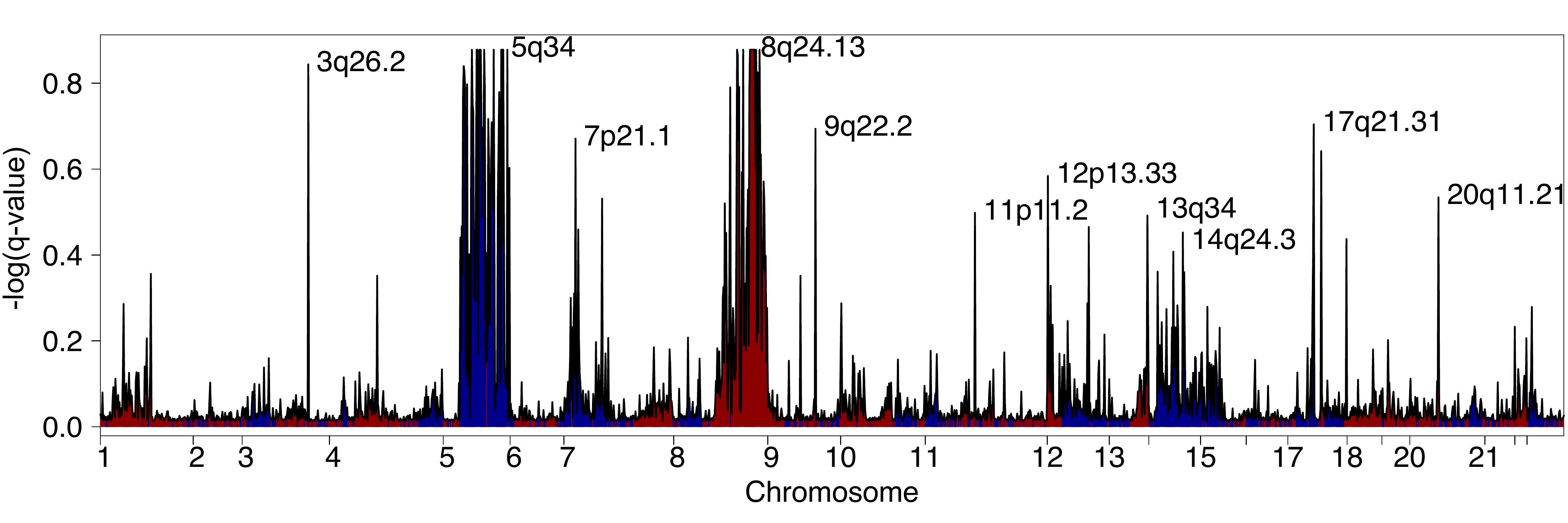}}

\smallskip
\textbf{B.} Lasso analysis of  {\it MYC} expression with genome-wide copy number data

\centerline{ \includegraphics[width=.5\textwidth]{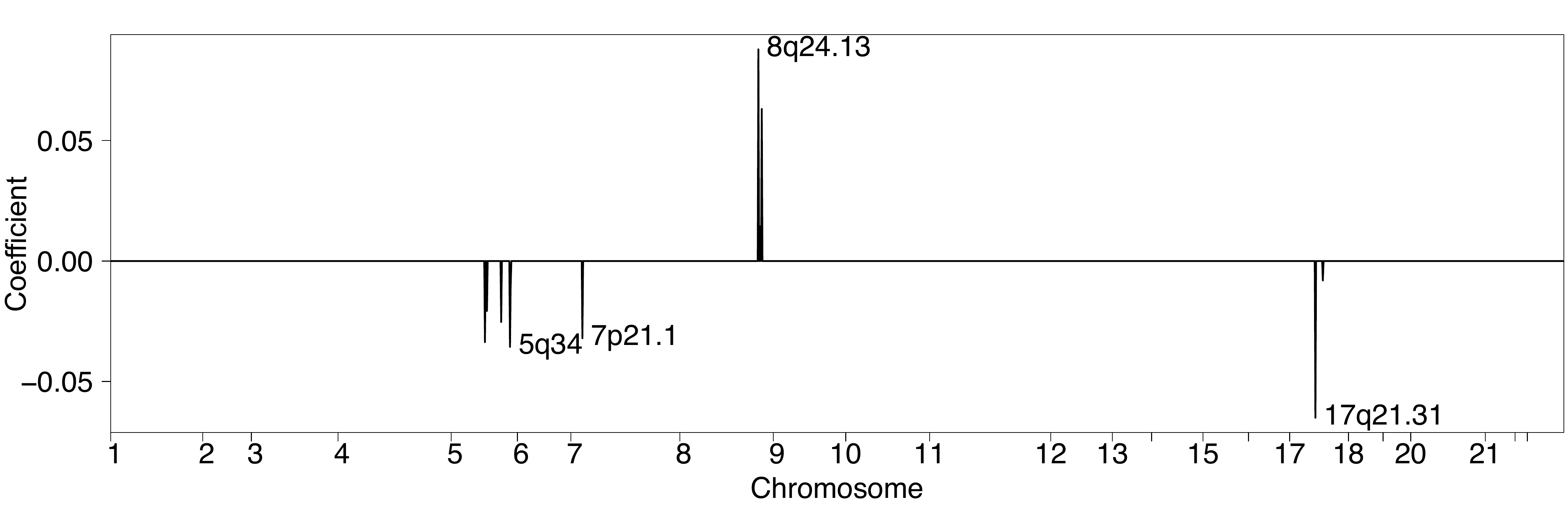}}
\caption{Example1:  {\it MYC}. The upper plot shows  the $q$-value (a multiple testing corrected $p$-value) of correlation coefficients between copy number and gene expression (red for positive correlations, blue for negative correlations). The lower plot shows the much sparser results from our framework, which illuminates key alterations by removing noise in the data.}
    \label{myc}
\end{figure}

We also examined the impact of copy number on individual target expressions. For the purpose of comparison, correlation of gene expression and genome-wide copy number was performed, and the results after FDR (False Discovery Rate) correction under the Benjamini-Hochberg procedure \cite{benjamini95controlling} is plotted in Fig.\ref{myc}.

Fig.\ref{myc}(a) shows that there is very high correlation between the expression of the oncogene {\it MYC} and copy number 3q26.2, 5q12-5q34. In fact, these regions exhibit nearly as high a correlation as does 8q24 (which denotes the location of  {\it MYC}). Here, the red histogram indicates positive correlation, while the blue histogram indicates negative correlation. Fig.\ref{myc}(b) illustrates that the dependence of  {\it MYC} expression on copy number in 5q21-5q34, when conditioned on its own copy number (8q24), is much smaller than that of copy number in 8q24, and the dependence on 3q26 vanishes. This is due to the fact that co-occurring events, such as co-amplification, have been removed by the conditional dependence assumption in Lasso. Moreover, the effects of copy number in 9q22.2 and 11p11.12 amongst others, shown in Fig.\ref{myc}(a) are completely removed when conditioning on the copy number at 8q24. In contrast, an apparently strong negative regulatory effect from 17q21.31, which harbours {\it WNK4} and the breast cancer susceptibility gene {\it BRCA1} persists, suggesting an interaction between these key genes.


\section{Discussion}


This paper presents an efficient framework for integrating genome-wide copy number and expression data. Both {\it cis}- and {\it trans}- regulatory effects from the genomic level to the transcriptomic level can be quantitatively measured using conditional probabilities. In particular, $L_1$-constrained regression is able to select relevant predictor variables to be included into the statistical model.  In summary, our proposed framework not only allows for a genome-wide description of the effect of gene dosage on gene expression, but also can partition copy-number dependent and independent transcriptional changes. While the former are useful to identify copy number and expression events co-ordinatedly deregulated during cancer progression, the latter are interesting as a starting point for validating putative oncogenes. Thus, our framework not only yields a global view of the interactions and suggests downstream targets of CNAs ripe for validation, it can also be used as a starting point for further focused analyses of the genomic basis of cancer. Our findings in the Chin dataset are currently being validated in another much larger breast cancer dataset, which has not yet been published. 



\bibliographystyle{IEEEtran}
\bibliography{genomic}

\end{document}